%% file: BBH_disk_lett.tex
\documentclass[twocolumn,superscriptaddress,showpacs,prl,aps,amsmath,amssymb,nofootinbib]{revtex4-1}                                        
\usepackage{natbib}
\usepackage{graphicx,color}
\usepackage{amsmath,amssymb}
\usepackage{verbatim}
\usepackage{float}
\usepackage{wasysym}
\usepackage{amssymb,graphicx}
\usepackage{epsfig}
\usepackage{psfrag}
\usepackage{dsfont}
\usepackage{amsfonts}
\usepackage{mathrsfs}
\usepackage{multirow}
\usepackage{times}
\usepackage{bm}
\usepackage{hyperref}
\usepackage{xspace}
\usepackage{rotating}
\hypersetup{
  colorlinks=true,        
  linkcolor=blue,         
  citecolor=cyan,         
}
\usepackage{pifont}
\usepackage[normalem]{ulem}   

\graphicspath{{./}}

\input kigou.tex

\newcommand{\GO}{\omega}

\newcommand{\be}{\begin{equation}}
\newcommand{\ee}{\end{equation}}

\def\QEQ{{%
    \setbox0\hbox{$I$}%
    \rlap{\hbox to \wd0{\hss--\hss}}\box0
}}

\begin{document}
\title{Multimessenger Signatures of Tilted, Self-Gravitating, Black Hole Disks}
\author{Milton Ruiz}
\affiliation{Departamento de Astronomía y Astrofísica, Universitat de València, Dr. Moliner 50, 46100, Burjassot (València), Spain}
\author{Antonios Tsokaros}
\affiliation{Department of Physics, University of Illinois at Urbana-Champaign, Urbana, IL 61801, USA}
\affiliation{National Center for Supercomputing Applications, University of Illinois Urbana-Champaign, Urbana, IL 61801, USA}
\affiliation{Research Center for Astronomy and Applied Mathematics, Academy of Athens, Athens 11527, Greece}
\author{Stuart L. Shapiro}
\affiliation{Department of Physics, University of Illinois at Urbana-Champaign, Urbana, IL 61801, USA}
\affiliation{Department of Astronomy, University of Illinois at Urbana-Champaign, Urbana, IL 61801, USA}
\affiliation{National Center for Supercomputing Applications, University of Illinois Urbana-Champaign, Urbana, IL 61801, USA}
\date{\today}
\begin{abstract}
We perform fully relativistic GRMHD simulations of magnetized, self-gravitating black hole–disk (BHD) systems 
in which the black hole spin is misaligned with the disk angular momentum. Massive disks ({disk to BH} mass ratios 
{of} $16-28\%$)  around rapidly rotating black holes ($\chi\lesssim 0.97$) develop a nonaxisymmetric instability 
for tilt angles from $0^\circ$ to $180^\circ$. Magnetic stresses damp, but {do} not completely {suppress}, the 
nonaxisymmetric instability, and {corresponding gravitational wave} (GW) emission, in aligned systems, while they 
enhance it in antialigned BHDs:  MRI-driven turbulence enhances angular momentum transport and accelerates nonlinear
{instability} evolution in misaligned  configurations. All models launch magnetically driven jets consistent with the 
Blandford–Znajek (BZ) mechanism,  with collimation depending on spin orientation. {The GWs reflect} strong nonaxisymmetric 
structure from  {a} persistent $m=1$ mode.  The coupling between fast MRI and the slower nonaxisymmetric instability
{growth} governs the outcome, with tilt controlling how MRI modifies the global mode. These simulations provide the first
self-consistent GRMHD treatment of tilted, self-gravitating BHD systems and support their role as multimessenger sources.
\end{abstract}
\maketitle
%
\textit{Introduction.}\textemdash
BHs across all mass scales are commonly surrounded by accretion disks formed in massive
star collapse~\cite{2009ARA&A..47...63S,OConnor:2010moj,Sun:2018gcl}, 
AGN activity~\cite{EventHorizonTelescope:2022wkp,EventHorizonTelescope:2022ago}, asymmetric 
supernovae~\cite{Gottlieb:2022tkb,Siegel:2017nub}, and compact object mergers~\cite{KAGRA:2021vkt}. 
These  systems are key multimessenger probes of strong gravity and high-energy 
astrophysics~\cite{Leong:2025qiw}. A crucial parameter is the BH spin, which can 
reach near-extreme values (e.g., $0.90\pm0.06$ for Sgr $\rm A^*$~\cite{Daly:2023axh}). When the 
BH spin is misaligned with the disk angular momentum, the system dynamics and observables can 
change substantially, affecting jet precession and power~\cite{Liska:2017alm,Jiang:2025xly}, accretion variability~\cite{Ingram:2019mna,Tsokaros:2022hjk}, and gravitational wave (GW) emission from 
non-axisymmetric disk dynamics~\cite{Tsokaros:2022hjk}.

The astrophysical relevance of spinning, misaligned BHs is highlighted by systems such as the blazar 
OJ 287~\cite{2026A&A...705A..23G}. While its quasi-periodic outbursts have often been interpreted as 
evidence for a supermassive BH binary, recent observations suggest alternative explanations involving 
a single misaligned BHD~\cite{Butuzova:2020khp}. Large spin–orbit misalignments may arise through several 
formation channels. In isolated binaries, natal supernova kicks can tilt the orbital plane~\cite{Hobbs:2005yx,
Kalogera:1999tq}. Gravitational wave (GW) observations indicate that a substantial fraction ($12$–$44\%$) 
of BH binaries exhibit spin–orbit misalignments larger than $90^\circ$~\cite{LIGOScientific:2020kqk}. Such 
processes can naturally produce tilted BHD systems, for example after BH–neutron star mergers in which 
the disrupted neutron star forms a misaligned accretion disk around the remnant BH~\cite{Foucart:2010eq}. 
The resulting geometry can strongly affect accretion, magnetic structure, jet stability, 
and the associated electromagnetic (EM) and GW signals.

BHD systems power magnetically driven relativistic jets from 
short $\gamma$-ray bursts to AGN{s}~\cite{prs15,Ruiz:2016rai,2011IJMPD..20.1547R}. In the standard picture, 
magnetic flux supplied by the accretion flow threads the BH and extracts its rotational energy to drive 
collimated outflows. Recent GRMHD simulations~\cite{Liska:2019vne,Liska:2017alm,Cui:2024ggx} suggest 
that strong BH spin–disk misalignment can modify this mechanism. Jet formation depends on the accumulation 
of large-scale poloidal flux and its coupling to a tilted, precessing disk~\cite{0264-9381-27-8-084013,Etienne:2012te,Etienne:2013qia,Ruiz:2020via,Cui:2024ggx}. Large misalignments 
can induce jet–disk interactions, magnetic 
flux eruptions, and intermittent outflows. Previous studies typically neglected disk self-gravity and considered
low-mass disks  in fixed spacetimes~\cite{Liska:2017alm,Liska:2019vne,2007ApJ...668..417F}. {By contrast, 
massive self-gravitating disks dynamically couple to the BH and modify the spacetime geometry. Although global 
instabilities such as the Papaloizou–Pringle instability (PPI) can arise even in non-self-gravitating tori, 
self-gravity can influence their growth and nonlinear evolution~\cite{Kiuchi2011b,Tsokaros:2022hjk,Wessel:2020hvu,
Wessel:2023jng}.} Understanding jet–disk interactions therefore requires a firm grasp of the underlying disk 
dynamics  in tilted systems.

The dynamics of tilted BHDs are {influenced} by the Lense–Thirring (LT) effect~\cite{1918PhyZ...19..156L}, in
which frame dragging from the spinning BH induces disk precession. In viscous disks this can lead to the 
Bardeen–Petterson (BP) effect~\cite{1975ApJ...195L..65B}, where the inner disk aligns with the BH spin. 
GR simulations of self-gravitating disks show instead that the disk and BH can undergo coupled precession 
and nutation, with LT torques driving global disk precession that in turn induces BH 
precession~\cite{Mewes:2015gma,Mewes:2016olz,Tsokaros:2022hjk}. Such disks may also develop the 
non-axisymmetric $m=1$ PPI~\cite{1984MNRAS.208..721P}, which redistributes angular momentum and produces a
persistent asymmetric mass distribution, making these systems potential sources of quasi-periodic GWs~\cite{Wessel:2020hvu,Shibata:2021sau,Tsokaros:2022hjk,Wessel:2023jng}.

Previous numerical studies have explored tilted disks in several regimes. Early GRMHD simulations of thick 
disks~\cite{Fragile:2004gp} found global precession without BP alignment and revealed plunging streams connecting 
the disk to the BH horizon. Later work on thinner disks identified disk tearing and standing shocks that enhance 
angular momentum transport at large tilt angles~\cite{Liska:2019vne,White:2019udt}. Ray-tracing 
calculations~\cite{2011ApJ...730...36D} further showed that the radiation edge of tilted disks can become largely 
independent of BH spin, while strongly magnetized disks may exhibit magnetically driven retrograde 
precession~\cite{Chatterjee:2023ber}. However, these studies neglected disk self-gravity and evolved low-mass 
disks in fixed spacetimes. GRMHD simulations of massive self-gravitating BHDs have mostly focused on aligned systems, 
where magnetic fields damp but do not suppress the PPI~\cite{Wessel:2023jng}. Highly spinning and tilted systems
have so far been explored only in hydrodynamic simulations~\cite{Mewes:2015gma,Mewes:2016olz,Tsokaros:2018zlf,Tsokaros:2022hjk}. 
A fully self-consistent GRMHD treatment of massive, tilted, self-gravitating BHDs is therefore still lacking.
We adopt units where $G=c=M_\odot =1$ throughout.

\begin{figure}                                                                   
\includegraphics[width=0.98\columnwidth]{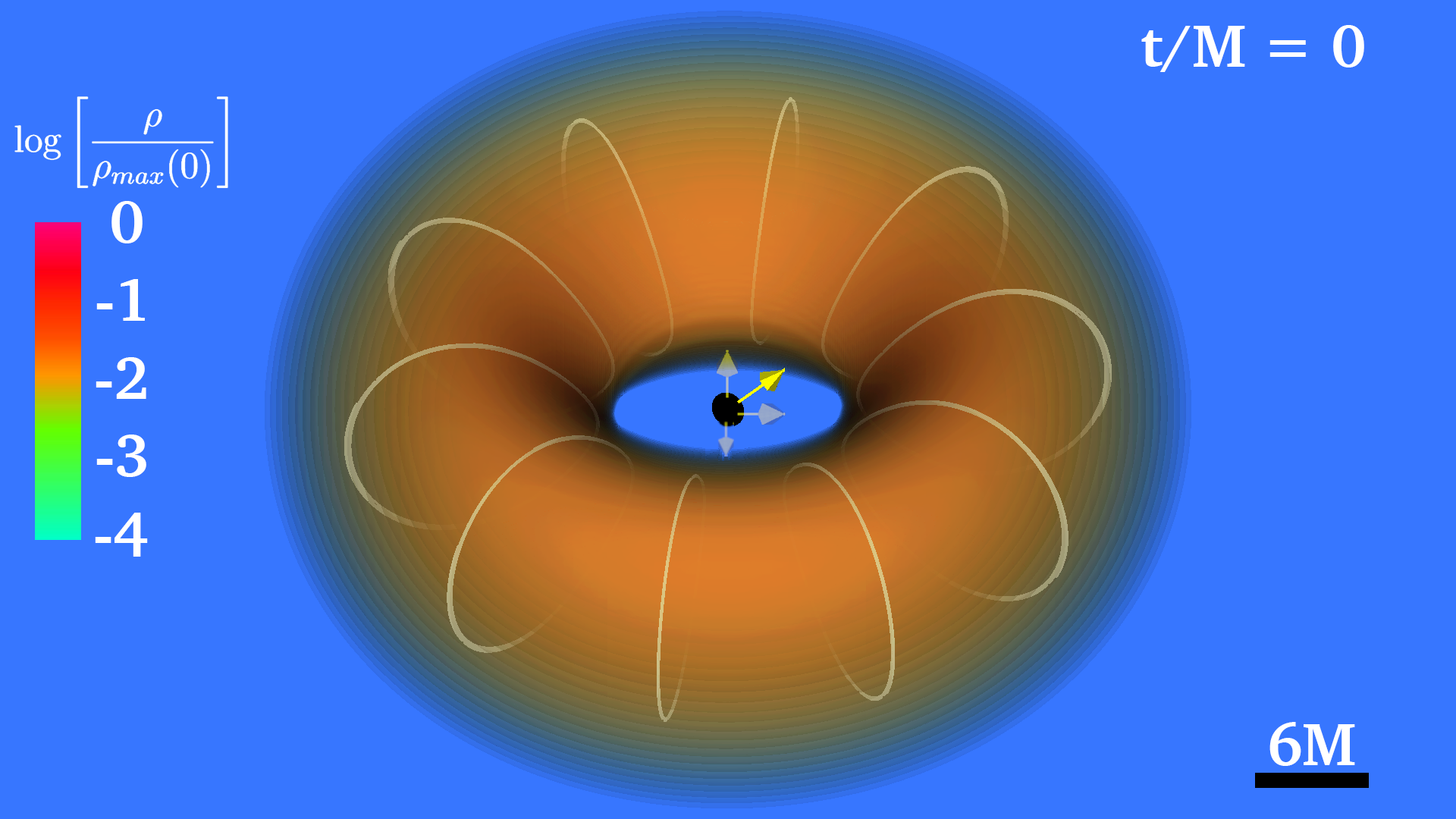}                        
\caption{ Volume rendering of the {initial} BHD system threaded by a purely poloidal magnetic 
field. The BH is depicted as a black spheroid. The yellow arrow shows the initial spin 
direction for case A2 $(45^\circ)$, while the fainter arrows indicate the spin orientations
for other cases (see Table I in~\cite{Tsokaros:2022hjk}).}  
\label{fig:ID_BNdisk}                                                                 
\end{figure}

%
\textit{Initial data and numerical methods.}\textemdash
We consider the self-gravitating BHD models (A1–A4) previously reported in~\cite{Tsokaros:2022hjk}. 
These equilibrium configurations were computed with the {\tt COCAL}  code using the methods 
of~\cite{Tsokaros:2018zlf}, which implement the Komatsu–Eriguchi–Hachisu scheme~\cite{1989MNRAS.237..355K} 
to construct initial data for massive, non-magnetized, self-gravitating tori. In all cases, the disk
angular momentum is initially aligned with the coordinate $z-$axis, while the near-extremal BH spin 
$\chi = J_{\rm BH}/M_{\rm BH}^2 \sim 0.97$ is tilted by angles~$\theta_s=0^\circ$, $45^\circ$, $90^\circ$, 
and $180^\circ$ (see the BH spin vectors in Fig. \ref{fig:ID_BNdisk}). The disk-to–BH mass ratios 
satisfy $M_0/M_{\rm BH}=0.156$–$0.280$ (see Table I in~\cite{Tsokaros:2022hjk}). The disk is described
by a $\Gamma= 4/3$ polytrope  with {nearly} constant  specific angular momentum ($j\sim r^{0.01}$). 
Here $M_0$ is the rest-mass of the disk, and~$J_{\rm BH}$ and $M_{\rm BH}$ are the quasi-local angular
momentum and the Christodoulou mass of  the  BH.

We {now} seed the disk with an {initially weak} poloidal magnetic field~(see Fig.
\ref{fig:ID_BNdisk}) using the same procedure as in~\cite{Ruiz:2023hit}. To ensure the resulting BHD is 
MRI-unstable, we set the initial magnetic-to-gas pressure ratio to $\beta^{-1}=0.01$. The self-gravitating 
BHD models A1–A4 are evolved using the {\tt ILLINOIS~GRMHD} code~\cite{Etienne:2010ui}. It employs the BSSN
formulation of Einstein equations, integrated with a fourth-order Runge-Kutta scheme and sixth-order finite
differencing. Moving mesh refinement is implemented via the~{\tt Carpet}  infrastructure~\cite{Schnetter:2003rb}. 
The hydrodynamics are evolved using high-resolution shock-capturing methods with an ideal gas $\Gamma-$law 
equation of state (EoS; $P=(\Gamma-1)\rho_0\,\epsilon$) with $\Gamma=4/3$ {which, e.g., would be the case for
a radiation dominated pressure support}. {The magnetic field evolution is treated within the ideal GRMHD 
approximation, assuming infinite conductivity and neglecting explicit resistive effects.}
Gauge conditions and constraint damping
are applied for numerical stability~\cite{Etienne:2011re}. All cases are evolved using 13 nested mesh refinement
boxes centered on the BH. The computational domain is $[-4000M_{\rm BH}, 4000M_{\rm BH}]^3$ with finest
resolution $\Delta x_{\rm min} = 50M_{\rm BH}/212 = 0.0122M_{BH}$. No symmetry assumptions are imposed.
%

\begin{figure*}                                                                   
\begin{turn}{90}
\hspace{1.7cm}\bf A1: Tilted $0^\circ$
\end{turn}
\includegraphics[width=0.98\columnwidth]{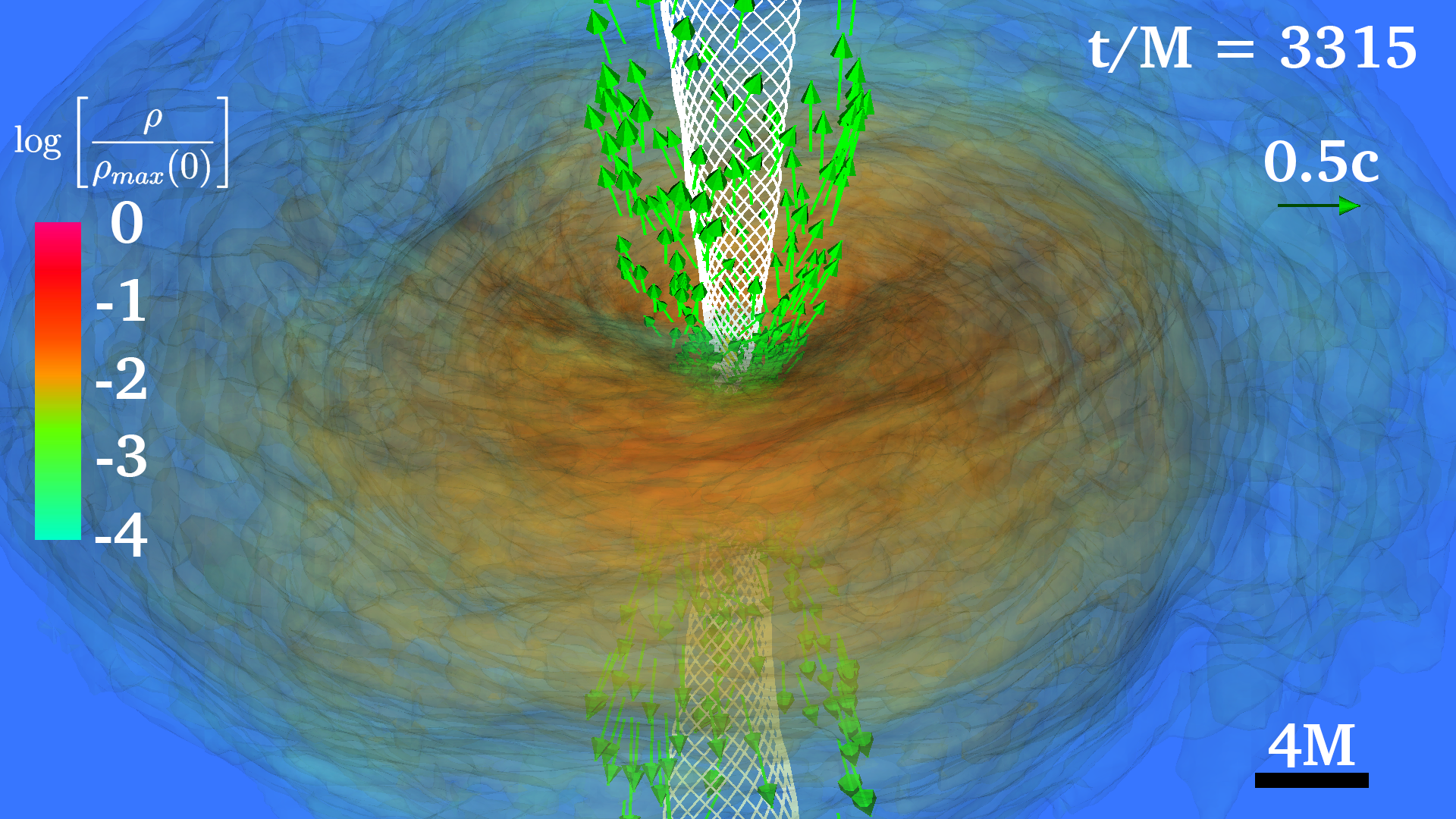}
\includegraphics[width=0.98\columnwidth]{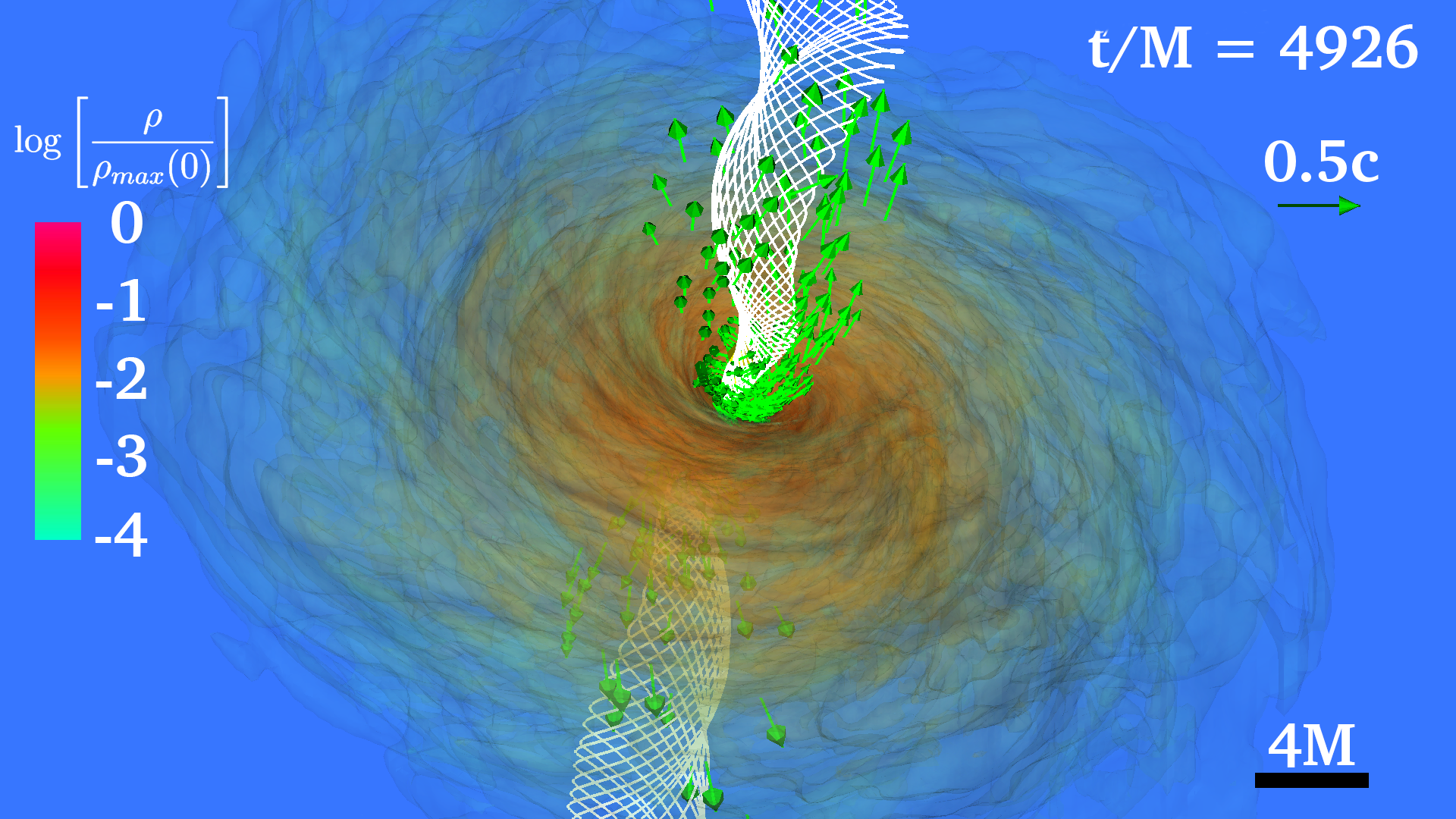}                            
\begin{turn}{90}
\hspace{1.7cm}\bf A2: Tilted $45^\circ$
\end{turn}

\begin{turn}{90}
\hspace{1.7cm} \bf A3: Tilted $90^\circ$
\end{turn}
\includegraphics[width=0.98\columnwidth]{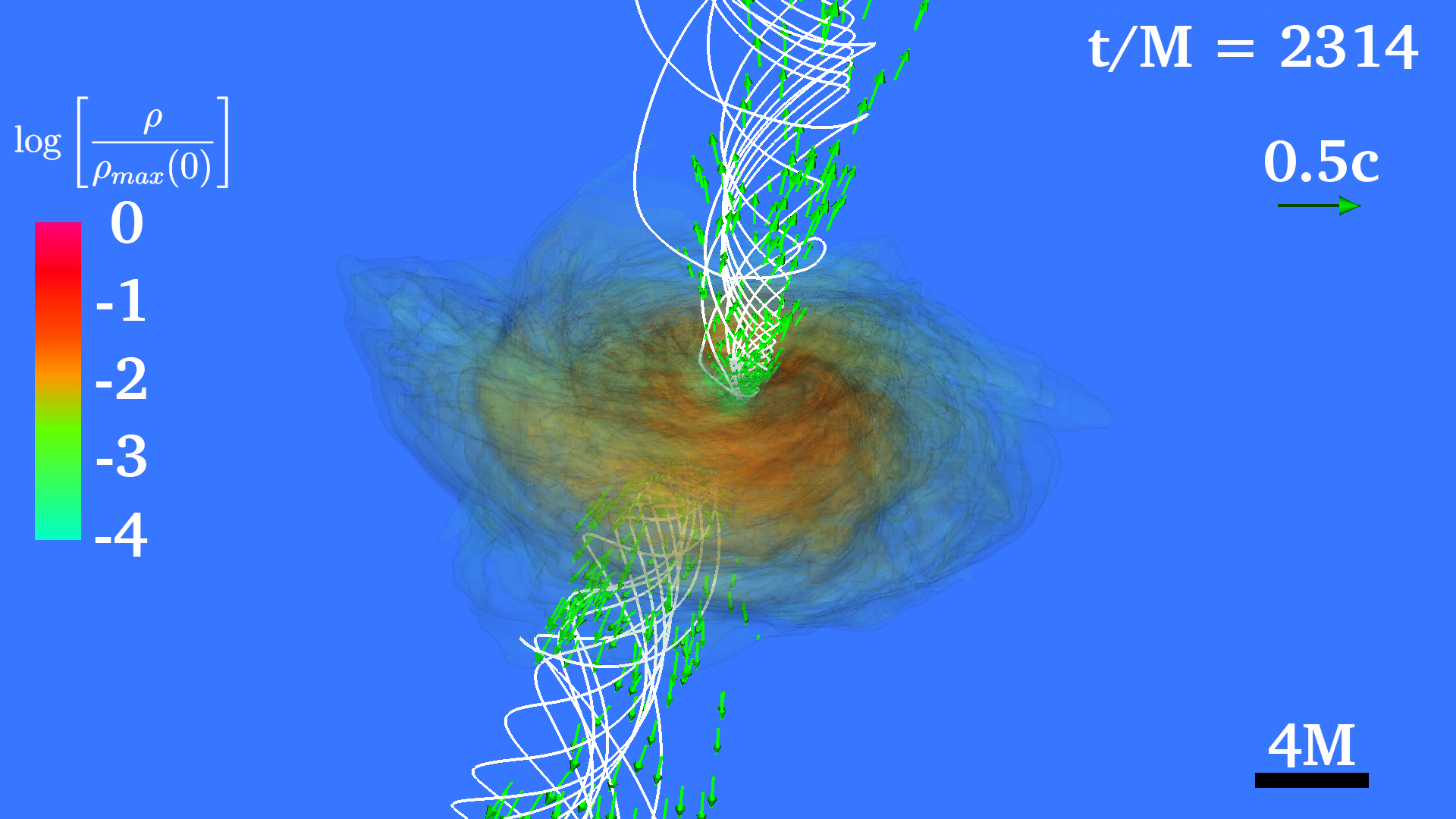}                            
\includegraphics[width=0.98\columnwidth]{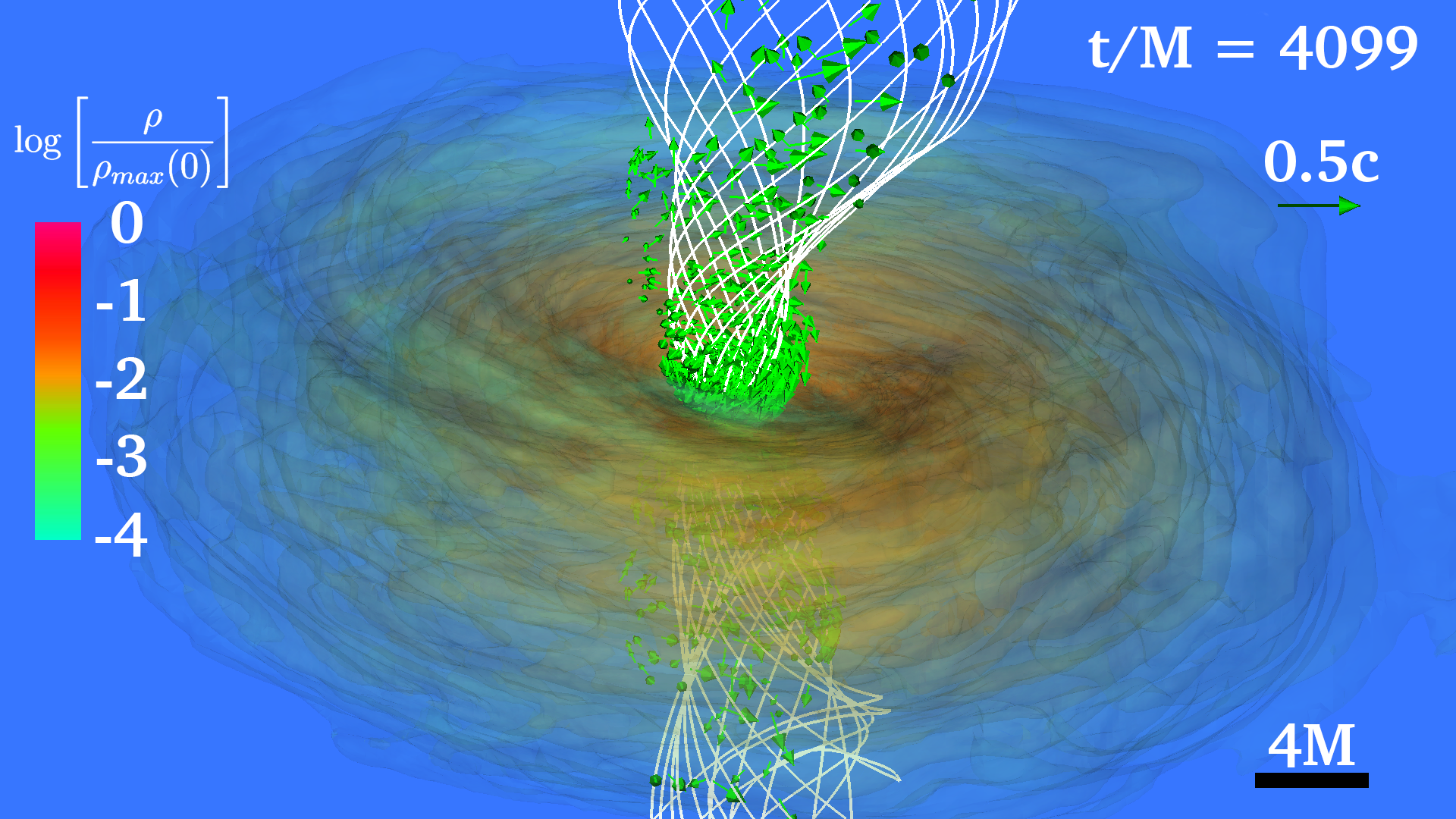}                            
\begin{turn}{90}
\hspace{1.7cm} \bf A4: Tilted $180^\circ$
\end{turn}
\caption{Volume rendering of the rest-mass density $\rho_0$ for the final outcome 
of BHDs (see Table I in~\cite{Tsokaros:2022hjk}), normalized to its initial maximum value 
(log scale). In all cases, a magnetically-driven, collimated outflow (jet) emerges along the 
BH polar regions. Arrows denote fluid velocities  while white lines trace the magnetic field,
illustrating its collimation into a jet-like structure. Here {the period} $P_c/M=\{370,404,357,422\}$.}  
\label{fig:final_outcome}                                                                 
\end{figure*}

\begin{figure}
\begin{center}
\includegraphics[width=0.98\columnwidth]{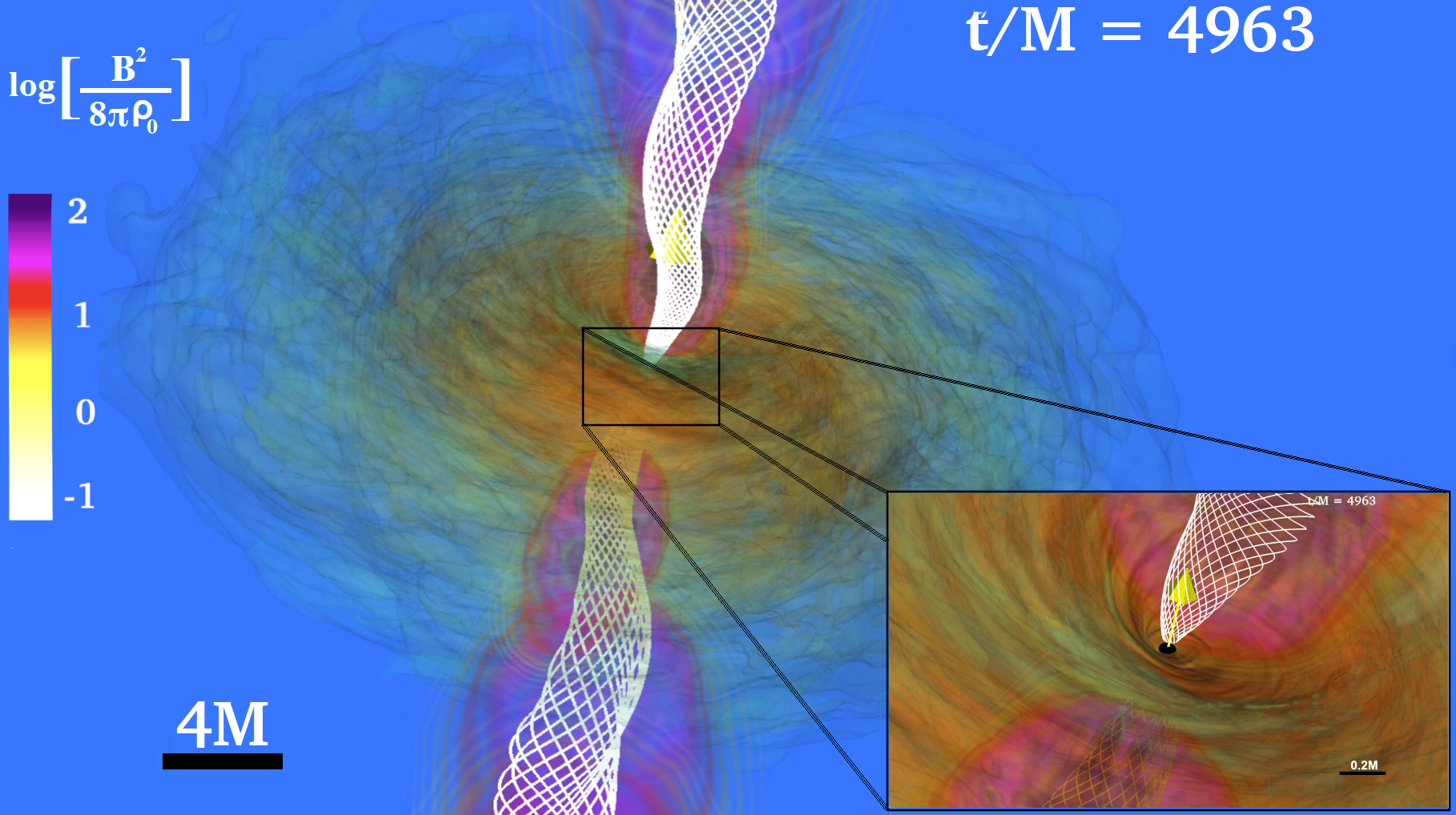}
\caption{Volume rendering of the rest-mass density $\rho_0$ in the disk ({color bar same as in Fig.~\ref{fig:final_outcome}}), 
and magnetization $B^2/(8\pi\rho_0)$ (log scale) in the funnel region near the end of simulation A2. White field lines mark regions
with $B^2/(8\pi\rho_0)\gtrsim10^{-2}$. The inset  shows horizon-threading field lines, magnetically dominated 
polar regions, and inner-disk alignment with the BH spin.}
\label{fig:A2_b2_over_2rho}
\end{center}
\end{figure}

\textit{Results.}\textemdash
The final outcome of the GRMHD simulations of models A1–A4 is shown in Fig.~\ref{fig:final_outcome}. 
In all cases the BHD launches a magnetically driven, collimated outflow from the BH poles, surrounded
by a turbulent, tilted disk whose magnetic field organizes into a jet-like structure. 
Fig.~\ref{fig:A2_b2_over_2rho} shows a representative tilted case (A2), highlighting 
magnetically-dominated regions and horizon-anchored field lines consistent with flux accumulation on the BH.
%
%

\begin{figure}                                                                                                                                    
\includegraphics[width=1.05\columnwidth]{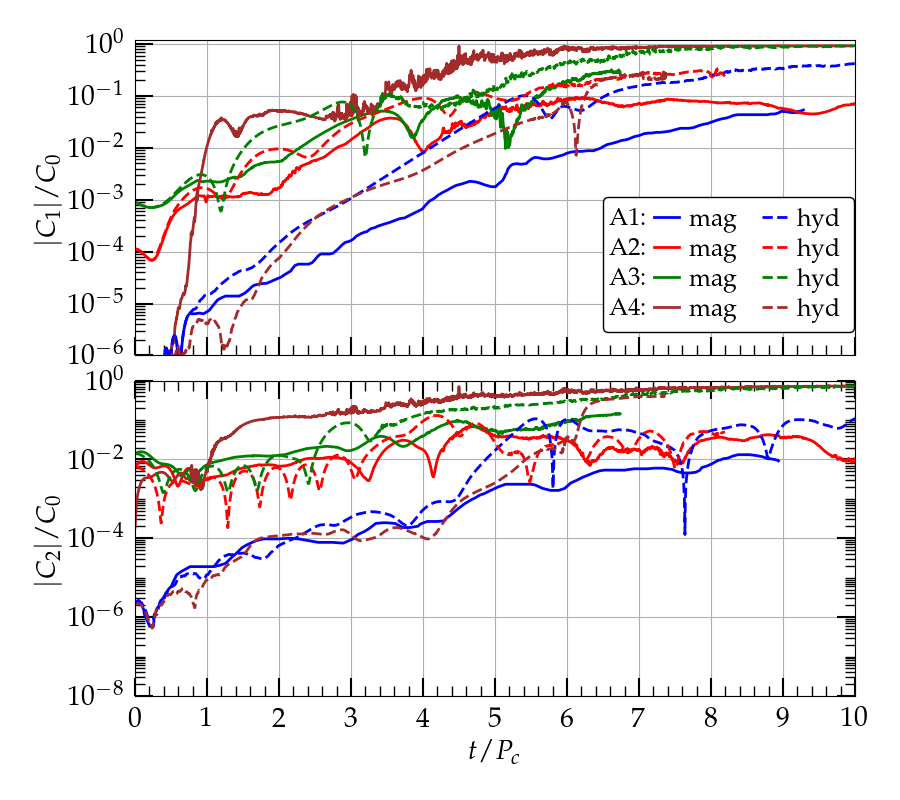}                        
\caption{Evolution of the amplitudes of the non axisymmetric $m=1$ and $m=2$ density modes, 
normalized by the $m=0$ mode. Solid lines denote the magnetized cases, while dashed lines indicate 
the {unmagnetized} cases reported in \cite{Tsokaros:2022hjk}.}  
\label{fig:modes}                                                                 
\end{figure}

These simulations represent the magnetized counterparts to the hydrodynamic models 
of~\cite{Tsokaros:2022hjk}. Fig.~\ref{fig:modes} compares the non-axisymmetric mode 
amplitudes $C_m=\int \rho_0 u^t\sqrt{-g}e^{im\phi}d^3x$~\cite{Wessel:2020hvu} for magnetized 
(solid) and hydrodynamic (dashed) evolutions. All configurations remain unstable to the PPI, 
with the antialigned case (A4) showing the most violent evolution, while the aligned case (A1) 
exhibits the weakest instability, with $m=1$ amplitudes about an order of magnitude smaller 
than in the hydrodynamic counterpart.

Magnetic turbulence is known to damp the $m=1$ mode in aligned BHD systems~\cite{Wessel:2023jng}. 
Unlike that study, where magnetization was introduced after PPI saturation, 
our simulations include magnetic fields from the start, allowing MRI and PPI to develop simultaneously. 
Significant damping occurs only in the aligned configuration (A1), where the growth rate is $\sim30\%$ 
lower than in the hydrodynamic model (Table~\ref{tab:mode_growth}). This weaker suppression suggests 
that coherent hydrodynamic structures can emerge from MRI-turbulent flows and dominate the dynamics, 
whether through standing shocks in tilted disks~\cite{White:2019udt} or through the PPI in our simulations.
\setlength{\tabcolsep}{5pt}
\begin{table}
\caption{Mode growth and pattern speed for the $m=1$ mode~\cite{Tsokaros:2022hjk}.}
\label{tab:mode_growth}
\begin{tabular}{cccccccccccccc}
\hline\hline
Model & ${\rm Im}(\GO_1)/\Omega_c$  & ${\rm Im}(\GO_1)/\Omega_c$ & $\Omega_{p,1}/\Omega_c$ & $\Omega_{p,1}/\Omega_c$ \\ 
      &    hyd                      &     mag                    &             hyd          &    mag       \\ \hline\hline
A1    & $0.318$                     &0.227                       & $0.748$                 &  $0.448$   \\ \hline
A2    & $0.177$                     &0.265                       & $0.748$                 &  $0.398$   \\ \hline
A3    & $0.177$                     &0.265                       & $0.637$                 &  $0.374$   \\ \hline
A4    & $0.227$                     &3.183                       & $0.812$                 &  $0.048$   \\ \hline
\end{tabular}                                                                                                                                            \end{table} 

In tilted configurations the PPI growth depends strongly on BH spin orientation: magnetized misaligned 
cases show a $\gtrsim50\%$ enhancement relative to the hydrodynamic models. As in~\cite{Tsokaros:2022hjk}, 
the antialigned case (A4) is the most unstable. This contrasts with thin, non-self-gravitating tilted 
disks~\cite{Liska:2019vne}, where large tilts produce disk tearing rather than stronger global modes. 
In our self-gravitating disks the PPI dominates even at high tilts, indicating that disk mass plays a 
key role in determining the governing instability. The $\sim 10-${fold} increase in growth rate for magnetized 
A4 relative to its hydrodynamic counterpart further shows that MRI-driven transport combined with tilt 
destabilizes the disk more effectively than either effect alone.

Magnetic turbulence triggers massive accretion on a timescale of $\sim 1\,P_c$ (see Supplemental Material), 
compared with $\sim 6\,P_c$ in the hydrodynamic case. {Here $P_c$ is the orbital period at the disk
radius of maximum rest-mass density}. In magnetized A4 the growth rate is amplified by a 
factor of $\sim 10$ (Table~\ref{tab:mode_growth}), producing rapid BH mass increase and spin-down to 
$\sim 0.5$ within $\sim 1\,P_c$. The hierarchy between MRI and PPI timescales explains this behavior. 
Our disks have nearly constant specific angular momentum ($j\sim r^{0.01}$), implying a shear parameter 
$q\sim2$. The MRI growth rate $\sigma_{\rm MRI}\sim\Omega_c$ yields $\tau_{\rm MRI}\approx 0.15P_c$, 
much shorter than the hydrodynamic PPI timescale ($\tau_{\rm PPI}\sim0.5$–$0.9P_c$)~\cite{Tsokaros:2022hjk}. 
MRI turbulence develops first, creating an effectively viscous background. In misaligned systems 
this turbulence enhances angular momentum transport and accelerates PPI growth, leading to stronger disk 
compression and earlier accretion.
%
%

The pattern speed $\Omega_{p,1}$ (Eq.~11 in~\cite{Tsokaros:2022hjk}) is systematically lower in the magnetized 
models (Table~\ref{tab:mode_growth}), and most pronounced in the antialigned case (A4) where it is $\sim 16$ 
times smaller than in the hydrodynamic counterpart. This reduction likely reflects MRI-driven angular momentum 
redistribution that alters the disk structure and shifts the characteristic radius of the global $m=1$ mode. 
The effect is strongest in tilted and antialigned systems where enhanced accretion and disk deformation 
accompany the instability.

\begin{figure}
\begin{center}
\includegraphics[width=1.0\columnwidth]{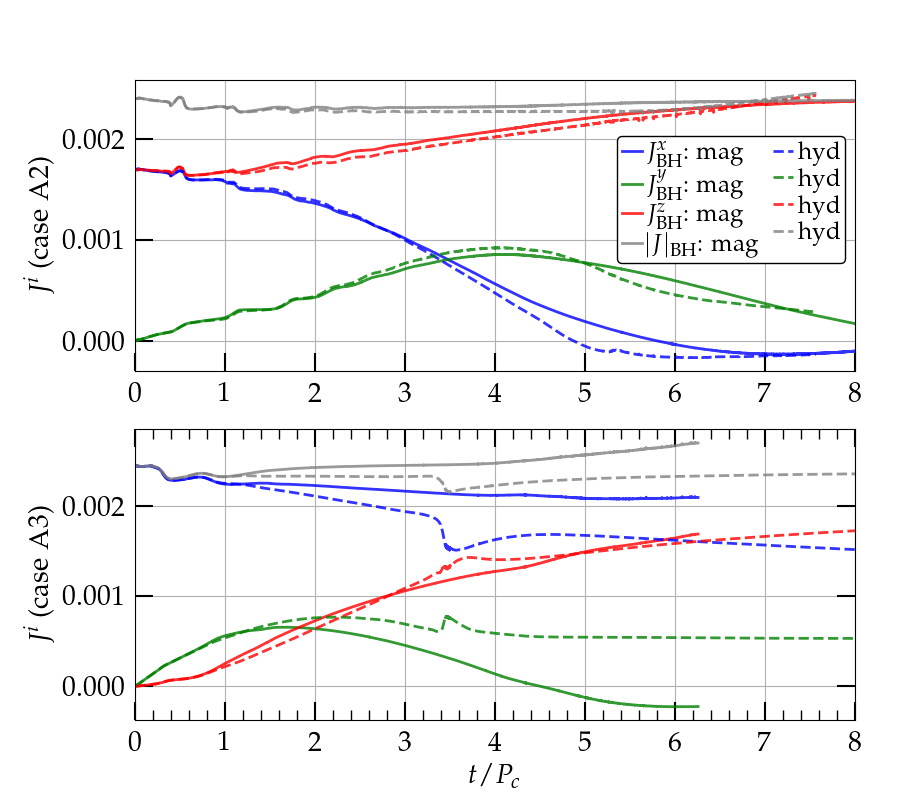}
\caption{Time evolution of the BH angular momentum components $J_i^{\rm BH}$ (colored) and 
magnitude $|J^{\rm BH}|$ (grey) for tilted cases A2 ($45^\circ$; top) and A3 ($90^\circ$; bottom).}
\label{fig:J_BH}
\end{center}
\end{figure}

Fig.~\ref{fig:J_BH} shows the evolution of the BH angular momentum components for 
tilted configurations A2 and A3. Magnetic stresses modify the nearly conservative 
precession seen in the hydrodynamic cases~\cite{Tsokaros:2022hjk}, introducing 
torque-driven exchange between the BH and disk. In A2 the magnetized evolution exhibits
smoother oscillations and gradual alignment of the spin with the disk angular momentum, 
while the spin magnitude remains nearly constant as $J_{\rm BH}$ and $M_{\rm BH}$ grow 
through accretion. This behavior reflects MRI-driven angular momentum redistribution and 
nonconservative BH–disk coupling. The gradual spin reorientation observed in A2–A3 differs 
from {non-self-gravitating} thick-disk simulations~\cite{2007ApJ...668..417F},  where global precession dominated 
with little alignment, likely reflecting the dynamical spacetime and stronger BHD  
angular-momentum exchange enabled by the self-gravitating disk in our models. The smoother 
evolution of magnetized simulations relative to hydrodynamic ones~\cite{Tsokaros:2022hjk} 
also contrasts with magnetically arrested disk (MAD) simulations~\cite{Chatterjee:2023ber}, 
where strong magnetic torques can drive retrograde precession that overwhelms LT  torques. 
Our models instead remain in the standard-and-normal-evolution (SANE) regime ($\beta\sim100$), 
where MRI-driven turbulence transports angular momentum but does not magnetically arrest 
the flow, leaving the dynamics LT-dominated. The rapid spin-down in A4 provides an alternative
pathway to similar outcomes: here angular-momentum cancellation during MRI-enhanced accretion
reduces the BH spin through a burst of accretion rather than sustained magnetic torques.

The differences are more pronounced in the $90^\circ$ configuration (A3). The hydrodynamic 
evolution shows an abrupt transition near $t\sim 3P_c$ associated with nonlinear PPI growth 
and a burst of accretion that reduces the spin to $\chi\sim0.85$. In contrast, the magnetized 
case evolves more smoothly: continuous angular momentum accretion gradually increases $J_{\rm BH}$ 
and produces a final spin $\chi\sim0.98$, while the spin vector slowly reorients toward the disk
axis (see Supplemental Material). This secular behavior is consistent with sustained MRI-driven
transport  and continuous accretion in tilted disks, which can gradually modify the BH spin magnitude 
and promote alignment with the disk angular momentum axis~\cite{2007ApJ...668..417F,Liska:2017alm}.

In the antialigned case (A4) magnetization produces qualitatively different behavior. MRI-driven 
turbulence triggers an early, intense accretion episode that rapidly reduces the BH spin to 
$\chi\sim0.5$ within $\sim1P_c$, {without leading to a spin-flip}. Here the spin evolution is 
dominated by angular momentum cancellation  during accretion rather than by gradual alignment or 
precessional dynamics.
%
%

The polar funnel forms as differential rotation winds the initially poloidal magnetic field, 
producing a magnetically dominated region ($B^2/8\pi\rho_0\gtrsim1$) above the BH poles. The 
resulting field geometry and velocities (Figs.~\ref{fig:final_outcome}–\ref{fig:A2_b2_over_2rho}) 
show a collimated outflow consistent with GRMHD simulations of tilted disks in fixed Kerr 
backgrounds~\cite{Liska:2017alm,Liska:2019vne,2007ApJ...668..417F}. In our dynamical spacetime 
the jets remain aligned with the BH spin rather than differentially precessing with the disk. 
The morphology depends on spin orientation: A1 shows the cleanest collimation, A2–A3 broader funnels, 
and A4 the most disrupted structure due to strong MRI-driven accretion and PPI activity.  
Collimated outflows persist in all configurations, showing that even highly tilted self-gravitating
disks can accumulate horizon-threading flux and launch BZ jets,  making jet formation a robust
outcome of BHD systems across tilt angles and for  disk-to-BH mass ratios up to $\sim 30\%$.

\begin{figure}
\begin{center}
\includegraphics[width=1.0\columnwidth]{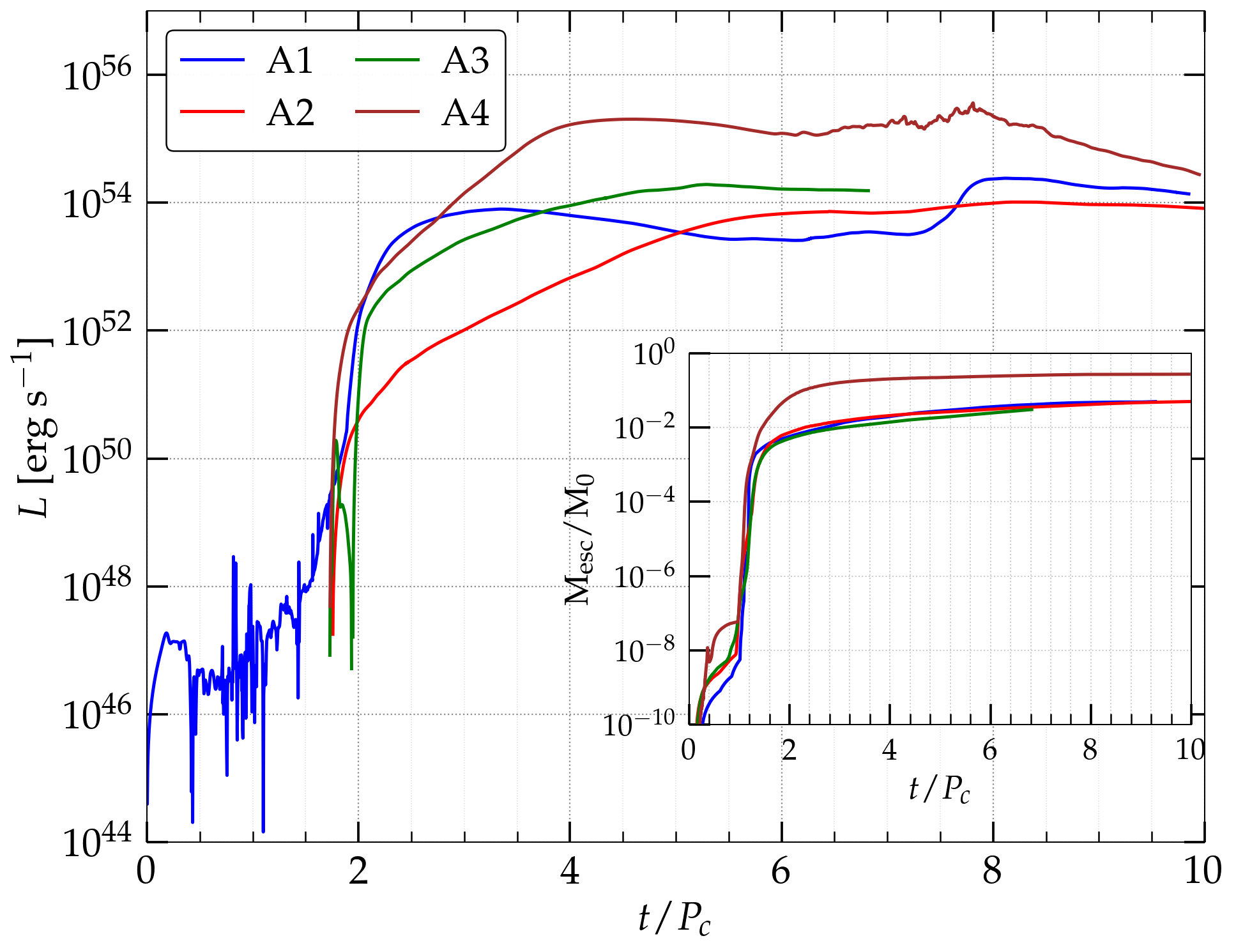}
\caption{EM Poynting luminosity $L_{\rm EM}$ as a function of time normalized 
to the orbital period $P_c$ for the  magnetized disks. The inset shows 
the corresponding fraction of unbound (escaping) mass.}
\label{fig:PLuinosity}
\end{center}
\end{figure}

The EM luminosity is $\sim10^{54}\,\rm erg\,s^{-1}$ in all cases (Fig.~\ref{fig:PLuinosity}), 
consistent with the BZ scaling $L_{\rm EM}\sim 10^{54}\chi_{0.9}^2M_{10^6}^2B_{10^{10}}^2\,{\rm erg
\,s^{-1}}$~\cite{1982MNRAS.199..883B}. The antialigned case exhibits the earliest and strongest peak, 
coinciding with the rapid accretion phase that drives substantial spin-down.  This behavior reflects efficient
inward advection of magnetic flux during the MRI-enhanced accretion episode.  By contrast, the aligned 
and moderately tilted cases maintain more stable luminosities consistent with their smoother accretion 
histories.
%
%
The inset of Fig.~\ref{fig:PLuinosity} shows that the dynamical ejecta masses are similar in most 
cases ($\gtrsim0.03M_0$), but increase by a factor of $\sim8$ in the antialigned configuration. This 
enhancement reflects the stronger accretion-driven outflows triggered by MRI turbulence. For 
stellar-mass BHs such ejecta ($>0.01M_\odot$) could power kilonovae comparable to 
GW170817~\cite{Cowperthwaite:2017dyu}, while in supermassive systems they may produce transient EM 
signatures such as AGN-like flares or tidal disruption event afterglows~\cite{Horesh:2021gvp}.

\begin{figure*}
\begin{center}
\includegraphics[width=1.0\columnwidth]{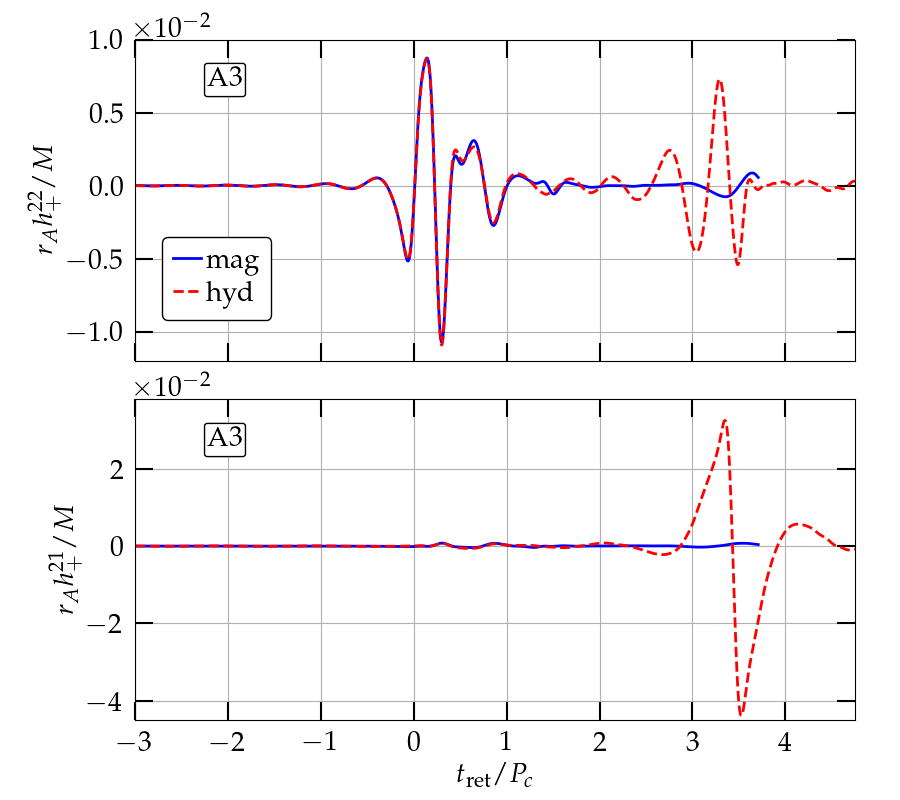}
\includegraphics[width=1.0\columnwidth]{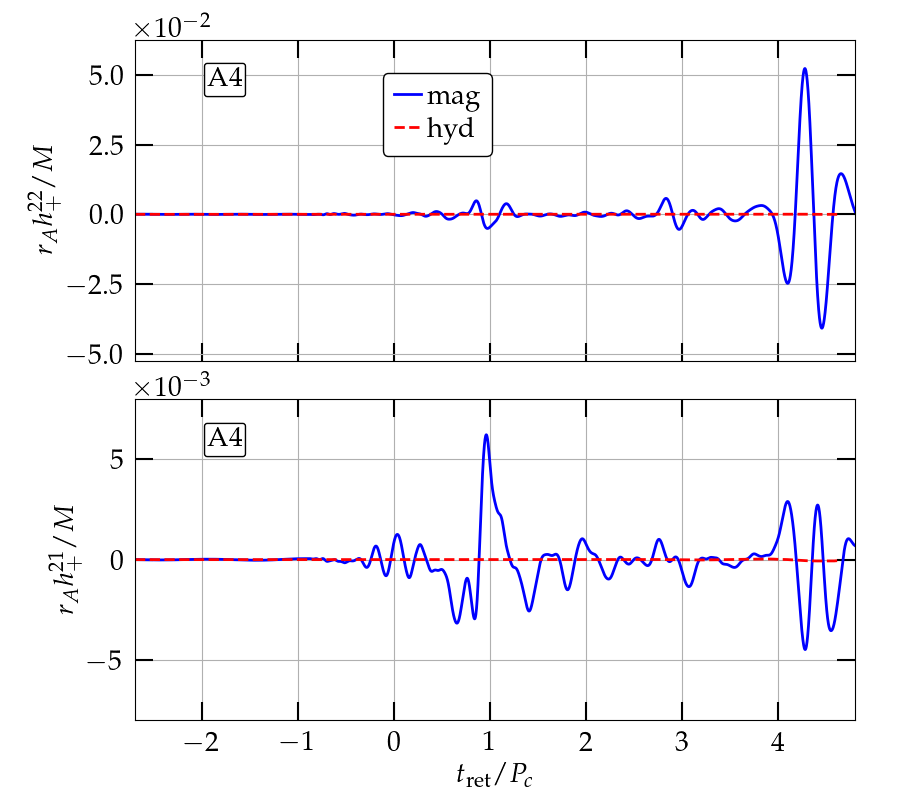}
\caption{EM
GW Strain $h+$ for {two} GW modes for the A3 (left) and A4 (right) cases.
Here $r_A$ is the areal extraction radius and $t_{\rm ret}$ is retarded time.}
\label{fig:GWamplitude}
\end{center}
\end{figure*}

Fig.~\ref{fig:GWamplitude} shows the GW strain for the $l=m=2$ (top) and $l=2,m=1$ (bottom) modes for 
representative tilted configurations A3 (left) and A4 (right), comparing magnetized evolutions (solid lines)
with their hydrodynamic counterparts (dashed lines). In aligned and moderately tilted systems ($\leq 90^\circ$), 
the magnetized waveforms have systematically smaller amplitudes, reflecting the damping of large-scale 
nonaxisymmetric modes by MRI-driven turbulence. However, in the antialigned case (A4), the magnetized evolution
produces larger amplitudes than the hydrodynamic case due to the violent instability and rapid MRI-triggered
accretion. This trend is also visible in the GW spectra (see Supplemental Material). Overall, magnetic fields
damp, but not completely suppress, GW emission in aligned systems but enhance it in antialigned BHDs.

\textit{Discussion.}\textemdash
Magnetic fields do not suppress the PPI in massive, self-gravitating, tilted BHDs. 
The instability persists in all configurations but develops earlier in misaligned 
systems, where MRI-driven turbulence enhances angular momentum transport. The tilt 
between the BH spin and disk angular momentum regulates the outcome: small tilts damp 
the nonaxisymmetric mode, near-orthogonal configurations ($\sim90^\circ$) recover 
amplitudes comparable to hydrodynamic evolutions, and antialigned systems ($>90^\circ$) 
strongly amplify it. In the $180^\circ$ case the $m=1$ growth rate increases by $\sim10$ 
relative to the hydrodynamic counterpart, triggering rapid accretion and spin-down. 
Magnetic stresses also introduce nonconservative BH–disk coupling, producing gradual 
spin reorientation in A2 and A3 and strong spin reduction only in the antialigned configuration.

All models launch magnetically dominated, horizon-threading outflows aligned with the BH
spin, consistent with the BZ mechanism. Misalignment does not prevent jet formation but 
affects collimation and variability, with the strongest early EM luminosity occurring in 
the antialigned case during rapid accretion. In contrast to fixed-background studies that 
neglect disk self-gravity~\cite{Liska:2017alm,Liska:2019vne,2007ApJ...668..417F}, the global
$m=1$ mode and disk warping in our simulations do not disrupt jet launching. The GW signal 
retains significant power beyond the $l=m=2$ mode. Magnetic turbulence generally reduces 
coherent GW amplitudes, except in the antialigned configuration where enhanced instability 
strengthens the signal.

{Because the GRMHD equations with a $\Gamma$-law EoS are scale free in geometrized
units, each simulation represents a family of BH+disk systems sharing the same
dimensionless parameters. The choice $\Gamma=4/3$ is adopted here as an effective 
description of the large-scale dynamics rather than a detailed microphysical EoS.
We note, however, that for supermassive BHs, the inner regions
of the disk are likely dominated by thermal radiation pressure,  for which
$\Gamma=4/3$ is a reasonable choice.}
{It is important to note that while very massive disks with $M_{\rm disk}/M_{\rm BH}\sim0.1$
may not be typical of standard AGN accretion flows, self-gravitating disks of this kind can
arise transiently in several astrophysical scenarios, including compact-object
mergers, collapsars, and some tidal disruption events (see e.g.~\cite{1993ApJ...405..273W,
  Goodman:2002gv,2020SSRv..216...35S}).
In addition, the formation scenario of supermassive BHs in the
early universe is unknown, and they could very well arise in gaseous regions whereby self-gravity
in the remnant disks is important and the pressure is radiation dominated. 
Our $\Gamma=4/3$ thermal EoS is appropriate for such disk systems.
More generally, the present work is intended to
isolate the fundamental nonlinear interplay among self-gravity, magnetic fields, and tilt
in relativistic BH+disk systems.}

%
%
%
%

%
\textit{Acknowledgments.}\textemdash
This work was supported in part by by
the Generalitat Valenciana Grant CIDEGENT/2021/046, 
by the Spanish Agencia Estatal de Investigación (grant PID2024-159689NB-C21) funded by MICIU/AEI/10.13039/501100011033 and by FEDER / EU,
the  National Science Foundation (NSF) Grants No. PHY-2308242 and No.  
OAC-2310548 to the University of Illinois at  Urbana-Champaign. 
A.T. acknowledges support from the National Center for Supercomputing Applications 
(NCSA) at the University of Illinois at Urbana-Champaign through the NCSA Fellows 
program. Further support has been provided by the  EU's Horizon 2020 Research and 
Innovation (RISE) programme H2020-MSCA-RISE-2017 (FunFiCO-777740) and by the EU 
Staff Ex-change (SE) programme HORIZON-MSCA-2021-SE-01 (NewFunFiCO-101086251). This 
work used Stampede2 at TACC and Anvil at Purdue University through allocation MCA99S008, 
from the Advanced Cyberinfras-tructure Coordination Ecosystem: Services \& Support
(ACCESS) program, which is supported by National Science Foundation grants \#2138259, 
\#2138286, \#2138307, \#2137603, and \#2138296. This research also used Frontera at TACC 
through allocation AST20025. Frontera is made possible by NSF award OAC-1818253. 
The authors thankfully acknowledge the computer resources at MareNostrum and the technical
support provided by the Barcelona Supercomputing Center (AECT-2025-3-0016 and AECT-2025-3-0017).   

\bibliographystyle{apsrev4-1}
\bibliography{references}
\clearpage
\input supplement.tex

\end{document}

%% file: kigou.tex
\newcommand{\beq}{\begin{equation}} 
\newcommand{\eeq}{\end{equation}} 
\newcommand{\beqn}{\begin{eqnarray}} 
\newcommand{\eeqn}{\end{eqnarray}}

\newcommand{\zD}{{\raise1.0ex\hbox{${}^{\ \circ}$}}\!\!\!\!\!D}
\newcommand{\alone}{{\raise0.5ex\hbox{${}^{\ 1}$}}\!\!\!\!\alpha}

\newcommand{\nalam}{\mathrel{\raise0.9ex\hbox{$^\lambda$}\mkern-14mu
\lower0.0ex\hbox{$\nabla$}}}

\newcommand{\zeroD}{{\raise1.0ex\hbox{${}^{\ \circ}$}}\!\!\!\!\!D}

\newcommand{\zLap}{{\raise1.0ex\hbox{${}^{\ \circ}$}}\!\!\!\!\Delta}
\newcommand{\zna}{{\raise1.0ex\hbox{${}^{\ \circ}$}}\!\!\!\!\!\nabla}
\newcommand{\zS}{{\raise1.0ex\hbox{${}^{\ \circ}$}}\!\!\!\!\!S}



%% file: supplement.tex
\section{Supplemental Material}
In this section, we further probe the dynamical stability of our BHDs.

\begin{figure}
\begin{center}
\includegraphics[width=1.0\columnwidth]{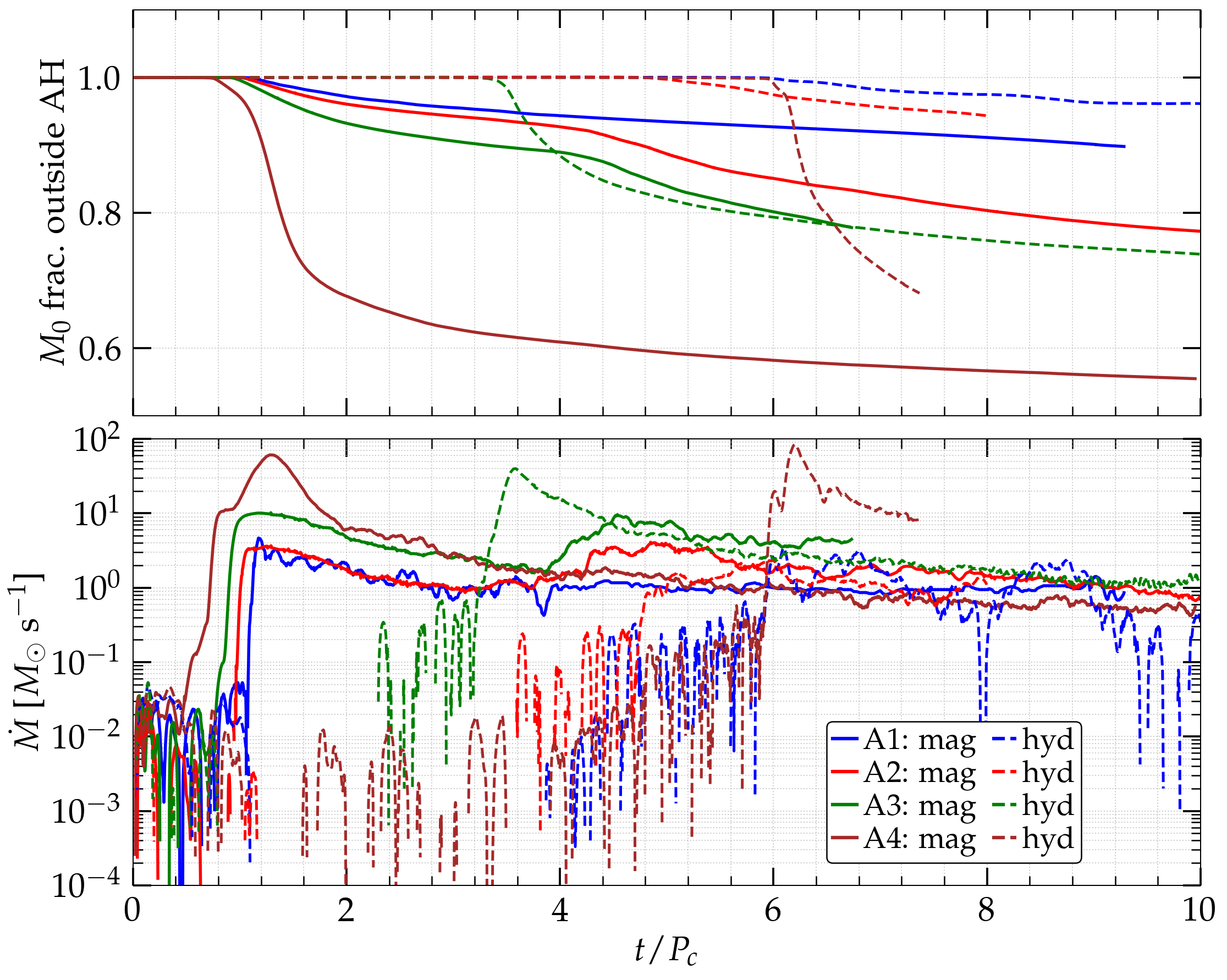}
\caption{Top: Rest-mass fraction remaining outside the apparent horizon as a function of 
time for all models. Bottom: Corresponding mass accretion history onto the black hole.}
\label{fig:fig_accretion}
\end{center}
\end{figure}

\textit{Accretion.}\textemdash
Fig.~\ref{fig:fig_accretion} shows the evolution of the rest-mass fraction outside
the apparent horizon and the corresponding accretion rate. The antialigned case (A4) 
exhibits the earliest and most dramatic mass depletion: by $\sim 1\,P_c$ the mass 
outside the horizon drops by nearly $30\%$, coincident with a sharp accretion peak. 
This early onset reflects the short MRI growth time ($\tau_{\rm MRI}\sim 0.15\,P_c$), 
which rapidly amplifies magnetic stresses. In A4, however,  the MRI interacts with 
strong non axisymmetric structure: the $m=1$ mode grows significantly faster than
in the hydrodynamic counterpart 
(see Fig.~4 in manuscript), {reaching large amplitudes within one orbital period and strongly distorting
the disk morphology}. The combined  action of MRI-driven turbulence and enhanced non 
axisymmetric dynamics produces the violent accretion episode.
%
%

This interpretation is supported by the evolution of the maximum rest-mass density 
(see Fig.~9 in manuscript).  In all magnetized cases $\rho_0^{\rm max}$ increases 
sharply by  $t\lesssim 1.5\,P_c$, 
indicating rapid compression as magnetic turbulence develops. The effect is most pronounced in
A4, where $\rho_{0,\max}$ rises by nearly an order of magnitude (reaching $\sim 16$ times its 
initial value) before dropping as accretion begins. Cases A1–A3 display more moderate 
amplification phases around $\sim 1.5\,P_c$  and $\sim 5\,P_c$, suggesting recurrent coupling 
between MRI turbulence and global PPI dynamics.
%
%
\begin{figure}
\begin{center}
\includegraphics[width=1.0\columnwidth]{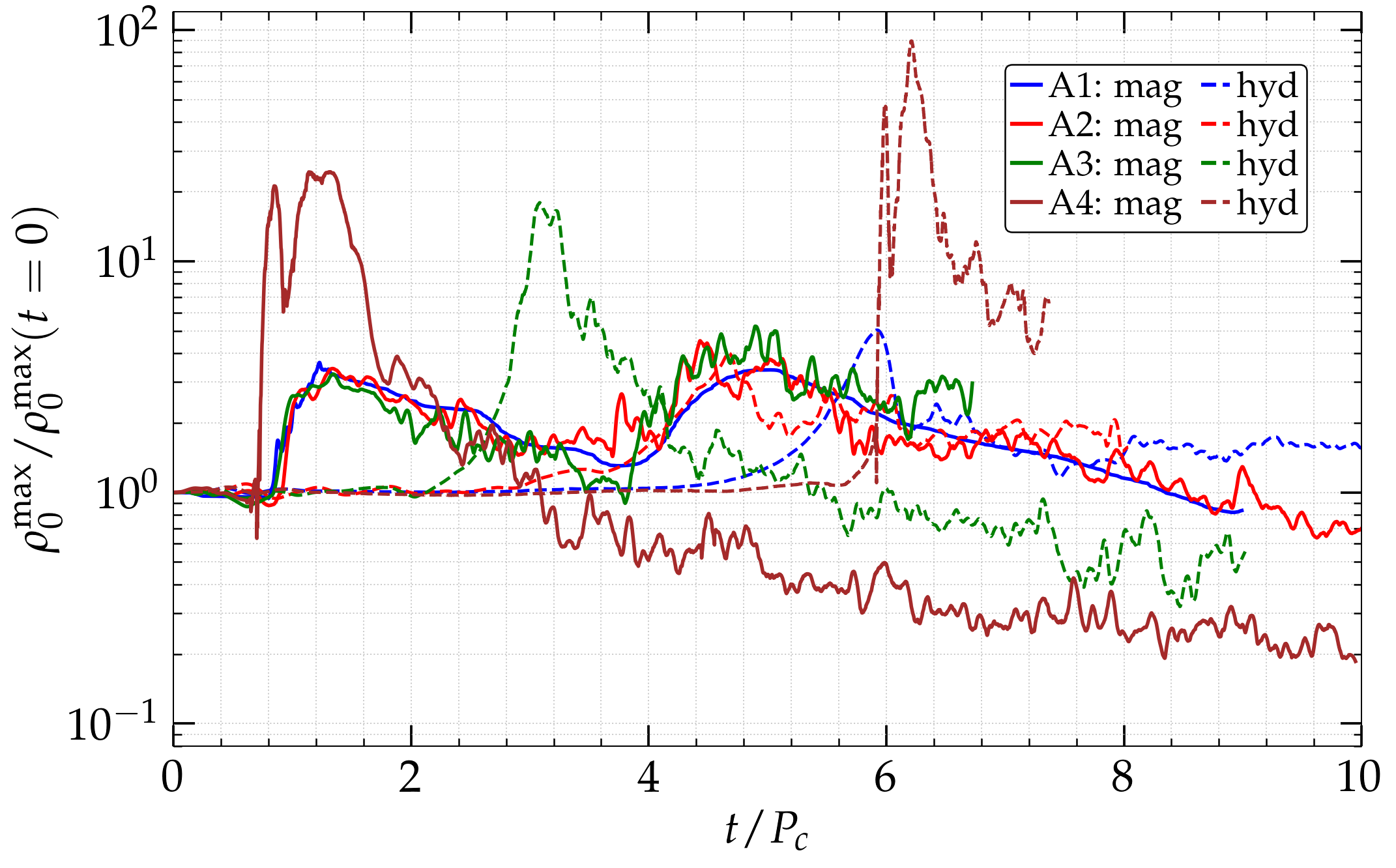}
\caption{Evolution of the maximum rest-mass density $\rho_0$ of the BHDs normalized by its
maximum value.}
\label{fig:rho_max}
\end{center}
\end{figure}

\begin{figure}
\begin{center}
\includegraphics[width=0.96\columnwidth]{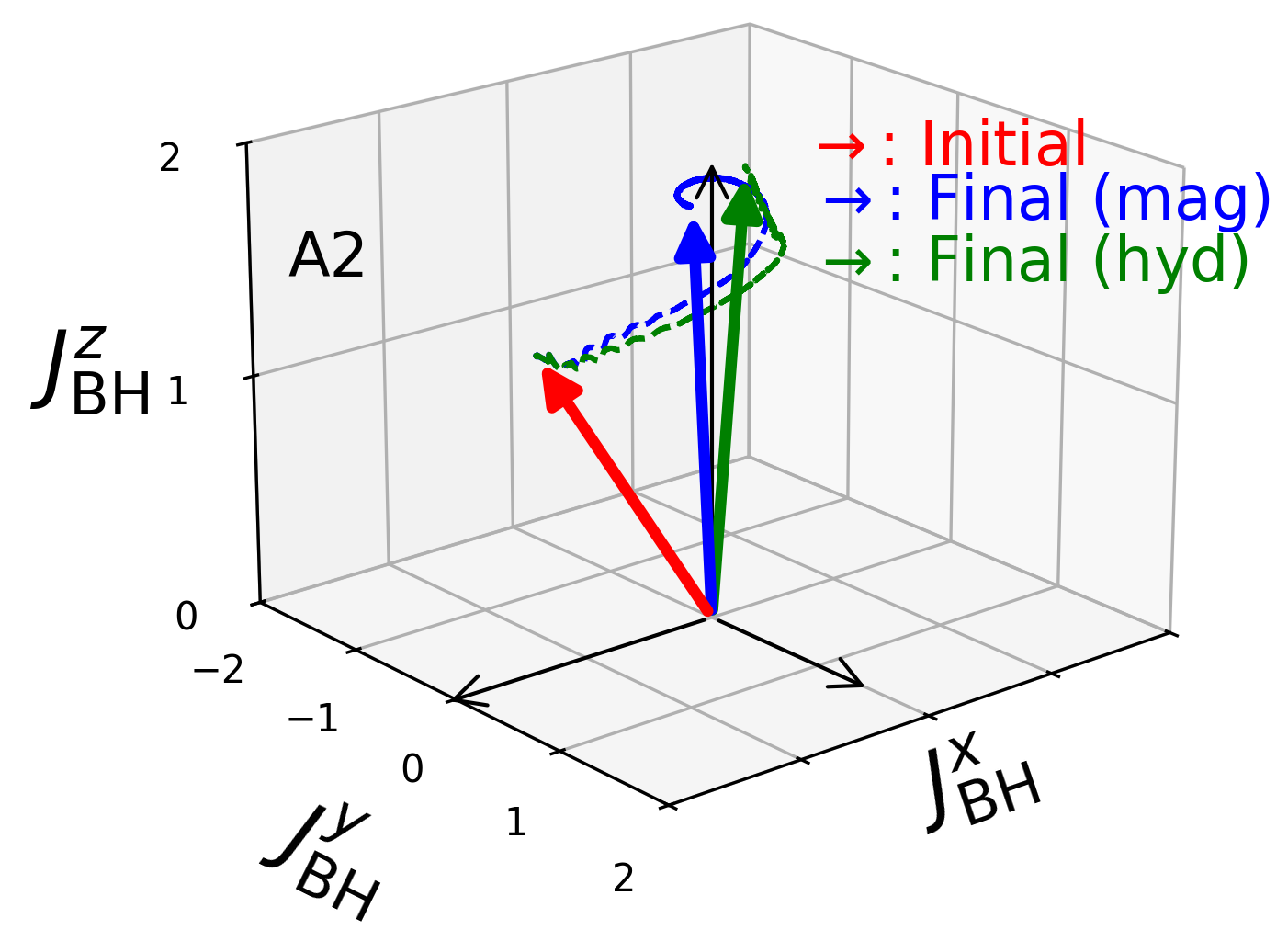}
\includegraphics[width=0.96\columnwidth]{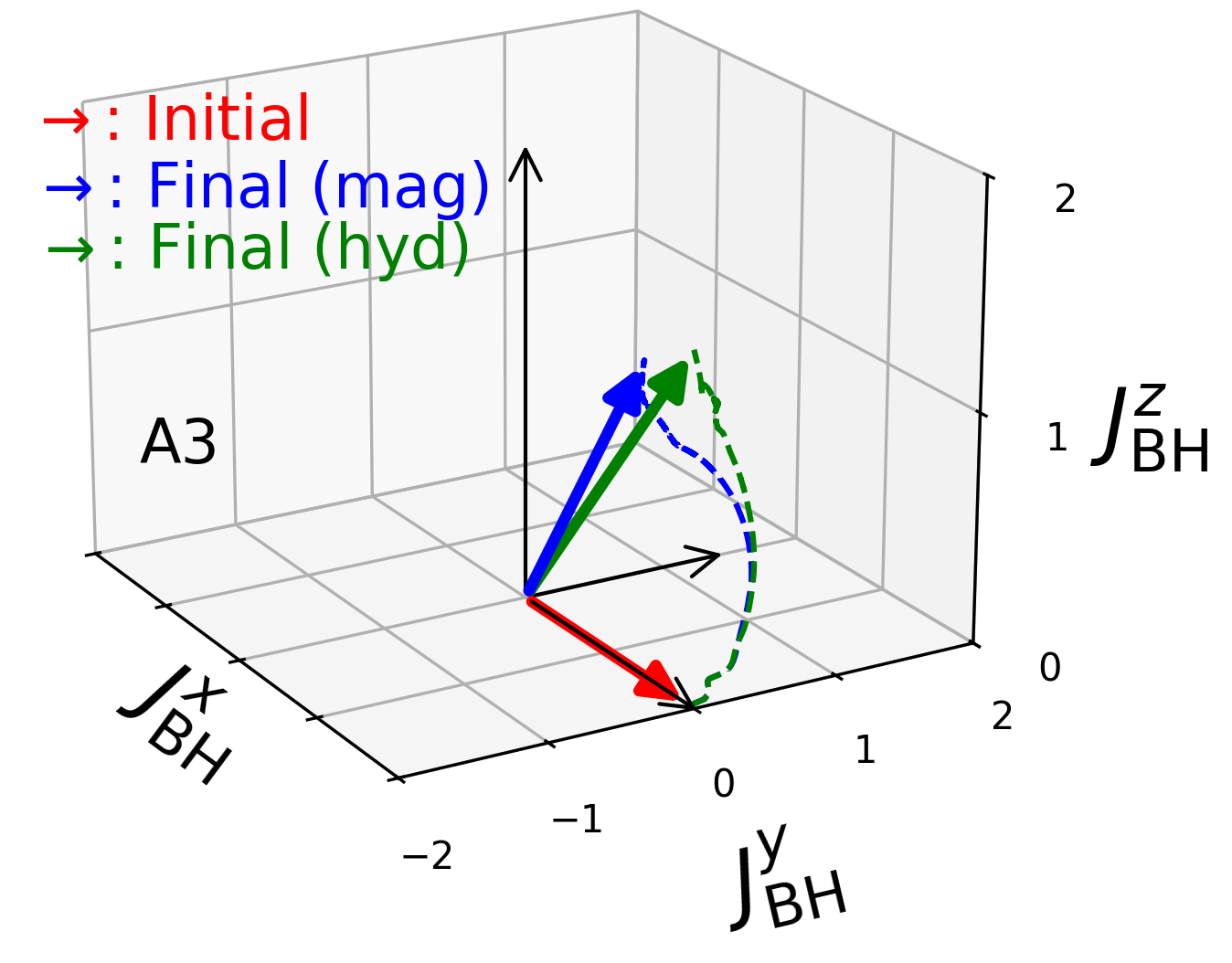}
\caption{ Comparison of the BH spin precession in the misaligned cases A2 at $t/P_c\sim 8.0$ (top) and 
A3 at $t/P_c\sim 6.5$~(bottom). The red arrow indicates the initial direction of the BH spin, while the 
blue and green arrows show its final direction in the magnetized and purely hydrodynamic cases, respectively. 
The dashed curve traces the evolution path of the spin vector. The spin magnitude is not to scale.}
\label{fig:spin_prec_pdep}
\end{center}
\end{figure}

\textit{Spin precession.}\textemdash
{Fig.~\ref{fig:spin_prec_pdep} compares the 3D evolution of the BH spin direction for 
both magnetized and hydrodynamic models in tilted configurations. For the $45^\circ$ case (A2), 
the trajectories in both cases exhibit similar precessional motion about the disk angular 
momentum axis, consistent with the LT precession~\cite{Tsokaros:2022hjk}. The magnetized
case shows only a modest reduction in oscillation amplitude over time, suggesting weak dissipative 
angular-momentum redistribution associated with magnetic stresses.} This behavior is consistent 
with a BP mechanism in which magnetic stresses promote alignment by coupling  the inner disk to the BH spin.

For the $90^\circ$ configuration (A3), both magnetized and hydrodynamic evolutions exhibit precession, 
but the magnetized trajectory evolves more smoothly and the spin magnitude increases modestly over time 
due to angular momentum accretion. This secular evolution reflects continuous coupling between the disk 
and the BH rather than a single dominant accretion episode. Although the geometric paths remain similar, 
magnetic stresses and sustained accretion modify the trajectory and promote a faster reorientation of 
the spin vector.

%
\textit{Gravitational wave spectrum.}\textemdash
\begin{figure}
\begin{center}
\includegraphics[width=1.0\columnwidth]{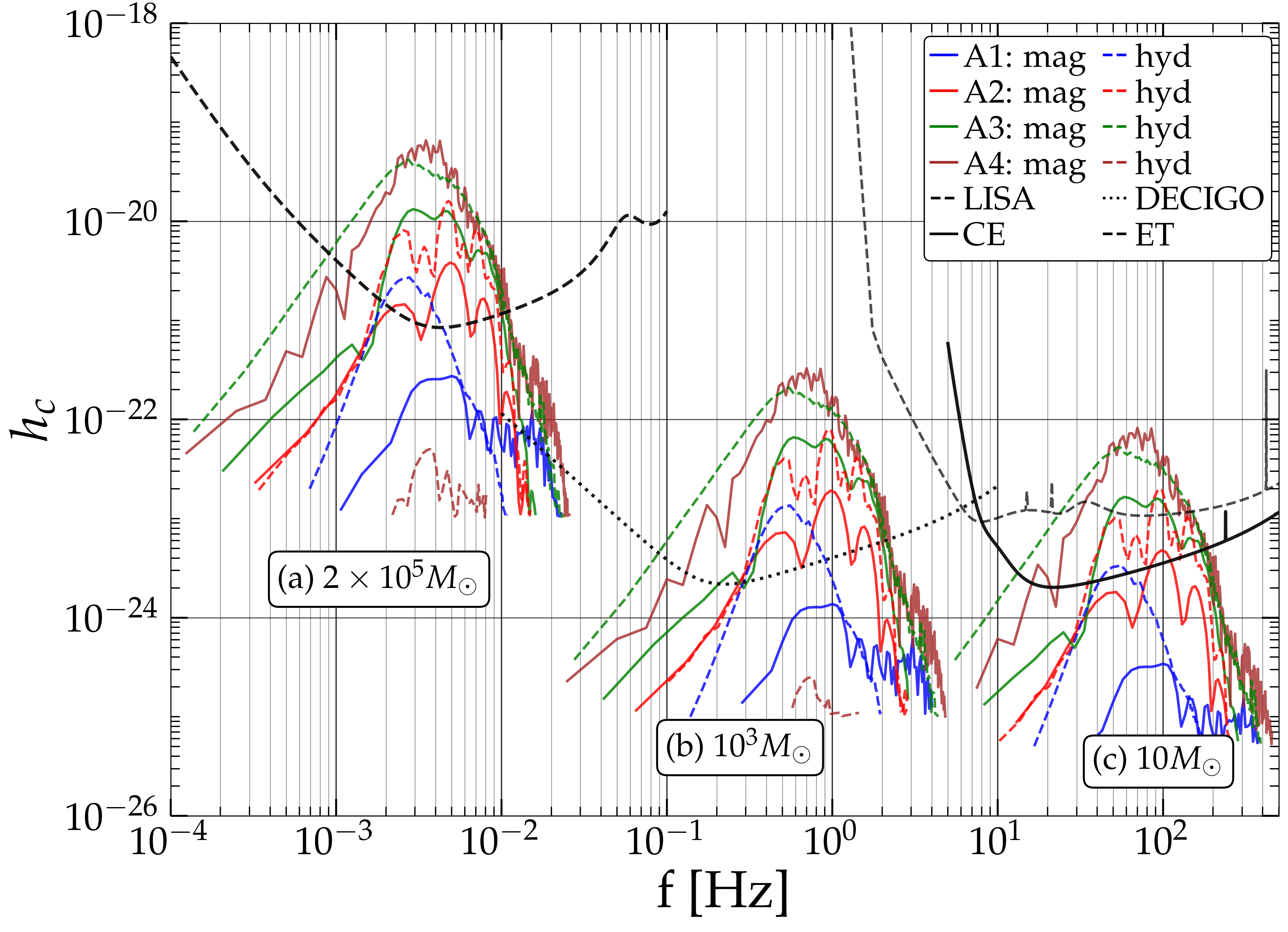}
\caption{ Characteristic strain of GW signals for magnetized (solid) and hydrodynamic (dashed) models,
shown for: (a) $2\times10^5\,M_\odot$ (LISA, $1000$ Mpc); (b) $10^3\,M_\odot$ (DECIGO, $1000$ Mpc); and (c)
$10\,M_\odot$  (CE/ET, $40$ Mpc). Curves are overlaid with detector sensitivities. Several models reach 
or exceed the sensitivity over a finite frequency range.}
\label{fig:PLuinosity}
\end{center}
\end{figure}
\begin{figure}
\begin{center}
\includegraphics[width=1.0\columnwidth]{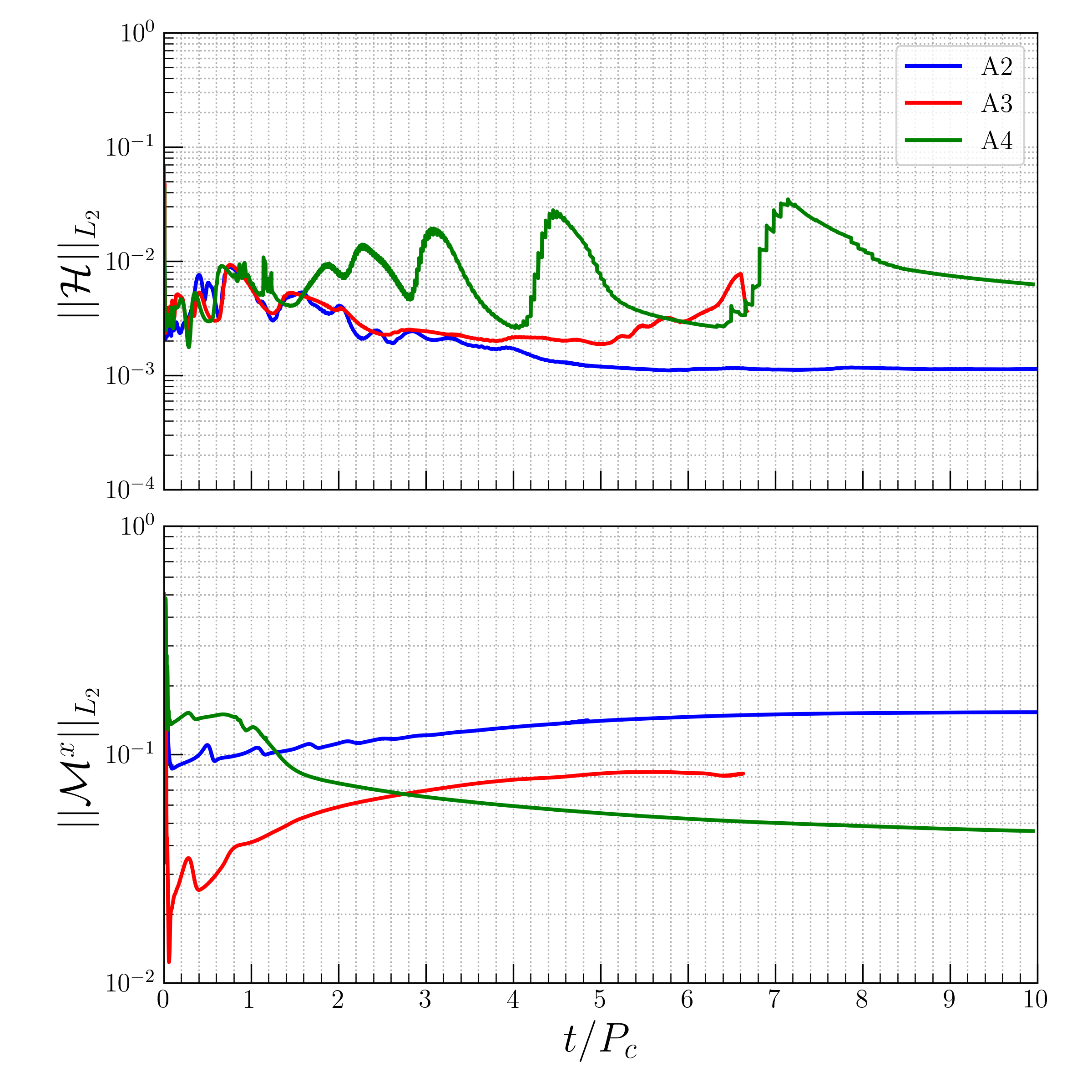}
\caption{Normalized violation of the Hamiltonian $\|H\|_{L_2}$
(top) and the x-component of the momentum constraint $\|\mathcal{M}^x|_{L_2}$ 
(bottom) .}
\label{fig:constraint}
\end{center}
\end{figure}
Fig.~\ref{fig:PLuinosity} shows the characteristic strain of the GW signals for the magnetized 
cases (continuous lines) and their purely hydrodynamic counterparts (dashed lines), scaled to 
representative mass ranges and distances and overlaid with the sensitivity curves of space- and 
third-generation ground-based detectors~\cite{Schmitz:2020syl}. The signals are shown for $2\times 
10^5\,M_\odot$ systems in the LISA band at $1000$~Mpc, $10^3\,M_\odot$ systems in the DECIGO band at 
$1000$~Mpc, and $10\,M_\odot$ systems in the CE/ET band at $40$~Mpc. A comparison with the detector
curves indicates that several models reach or exceed the {appropiate} sensitivity {curve} over a finite 
frequency range, while others remain below threshold, highlighting a strong dependence on the system 
configuration and on the development of dynamical and magnetized instabilities.

Consistent with~\cite{Wessel:2023jng}, the magnetized A1–A3 cases exhibit lower amplitudes than the 
hydrodynamic ones, highlighting the damping influence of magnetic turbulence on non-axisymmetric modes
(see~Fig.~4 in manuscript). 
By contrast, the antialigned case (A4) shows a higher peak amplitude in 
the magnetized simulation, driven by strong instability and enhanced accretion triggered by MRI-induced 
turbulence. This rapid accretion amplifies the GW emission, consistent with the stronger PPI growth 
observed in magnetized A4 
(see~Fig.~4 in manuscript). This behavior is consistent with the hydrodynamic
trends reported in~\cite{Tsokaros:2022hjk}, where tilt influences the GW emission, while here magnetic
fields modulate the amplitude through damping or enhancement depending on the alignment configuration. 
The $\sim 2$ higher amplitude in magnetized A4 relative to its hydrodynamic counterpart indicates that 
MRI-triggered accretion can boost GW emission beyond hydrodynamic predictions.

As in the hydrodynamic models, the characteristic strain peaks at approximately twice the orbital frequency
$f_c$ for A1 and A3, while A2 displays a primary peak near $3f_c$ with a secondary at $2f_c$. Our simulations
capture the early-time spectral structure of the signal, including amplitude and frequency modulations visible
for $t_{\rm ret} \lesssim 2 P_c$, associated with the initial nonlinear {instability} development of the disk. At later
times, these modulations diminish and the emission approaches a more quasimonochromatic signal, dominated 
by the persistent $m=1$ mode. These results indicate that BHD systems across a wide range of mass
scales may constitute promising targets for future GW observations, while magnetic fields play a key role 
in modulating their detectability by either suppressing or enhancing the GW emission depending on the alignment
configuration.

\textit{Constraint monitoring.}\textemdash
{To assess the accuracy and stability of the simulations, we monitored 
the evolution of the  $L_2$ norms of the normalized Hamiltonian and momentum 
constraints (x-component) for all our BH+disk case configurations  computed following Eqs. 
(40) and (41) in~\cite{eflstb08}. Fig.~\ref{fig:constraint} show their evolution
 for our tilted A2-A4 cases. In all cases the constraint remain bounded and below	
 $\lesssim 0.1$  throughout the simulations. The overall behavior is consistent with that 
 reported in previous long-term relativistic simulations (see e.g.~\cite{Tsokaros:2019anx,Ruiz:2021qmm}).}